\documentclass[a4paper,10pt,oneside]{article}
\usepackage[utf8]{inputenc}
\usepackage{icad2023,amsmath,epsfig,times,url,hyperref}
\usepackage[T1]{fontenc}
\usepackage{xspace}
\usepackage{microtype}
\usepackage{rotating}
\usepackage{booktabs}
\usepackage{threeparttable}
\usepackage{tabularx}
\usepackage{subfigure}
\usepackage{IEEEtrantools}

\newcommand{\project}[1]{\textsl{#1}\xspace}
\newcommand{\kepler}{\project{Kepler}}
\newcommand{\tess}{\project{TESS}}
\newcommand{\gaia}{\project{Gaia}}

\graphicspath{{./}{figures/}}

\title{The Sonified Hertzsprung-Russell Diagram}

\name{Daniela Huppenkothen$^{1}$\sthanks{\hspace{0.15cm}Corresponding author; d.huppenkothen@sron.nl}, Juan Pampin$^{2}$, James R. A. Davenport$^{3}$, James Wenlock$^{2}$ }

\address{$^{1}$SRON Netherlands Institute for Space Research, The Netherlands  \\
$^{2}$Department of Digital Arts and Experimental Media, University of Washington, Seattle, USA \\
$^{3}$Department of Astronomy, University of Washington, Seattle, USA}

\begin{document}
\bstctlcite{IEEEexample:BSTcontrol}
\ninept
\maketitle
\begin{sloppy}
\begin{abstract}
Understanding the physical properties of stars, and putting these properties into the context of stellar evolution, is a core challenge in astronomical research. A key visualization in studying stellar evolution is the Hertzsprung-Russell diagram (HRD), organizing data about stellar luminosity and colour into a form that is informative about stellar structure and evolution. However, connecting the HRD with other sources of information, including stellar time series, is an outstanding challenge. Here we present a new method to turn stellar time series into sound. This method encodes physically meaningful features such that auditory comparisons between sonifications of different stars preserve astrophysical differences between them. We present an interactive multimedia version of the HRD that combines both visual and auditory components and that allows exploration of different types of stars both on and off the main sequence through both visual and auditory media.
\end{abstract}

\section{Introduction}
\label{sec:intro}

There is perhaps no other visualization so central to both research and teaching of astronomy as the \textit{Hertzsprung-Russell diagram} (HRD), and its observational counterpart the colour--magnitude diagram. Familiar to any first-year astronomy student, it plots the optical colour of a star as a proxy for its temperature against its absolute magnitude as a measure of the star's intrinsic brightness. Even taken by itself, it encodes an astonishing amount of information for a single visualization: that most stars live on a narrow, diagonal band called the \textit{main-sequence}, that a population of red but luminous stars exist above this main sequence, and another population of small and hot stars below. With enough data, a second parallel track above the main sequence becomes visible made of binary systems with constituents of nearly equal masses \cite{gaiahrd}. HRDs have been used to explore the solar neighbourhood \cite{daltio2021}, individual globular clusters \cite{husser2016}, or the Milky Way as a whole \cite{gaiahrd}. 
Evolutionary tracks and isochrones derived from theoretical models of stellar evolution show how stars of varying metallicities move across the HRD onto the main sequence, evolve off the main sequence to become red giants, and some eventually into white dwarfs \cite{MIST}. In  this way we can use the HRD to study nearly every fundamental parameter of stars and stellar populations in our Galaxy.

The \gaia~Space Telescope has revolutionized stellar and galactic astrophysics in a number of ways. In particular, the release of nearly 2 billion stars with accurate distance estimates has yielded the most comprehensive colour--magnitude diagram ever created \cite{gaiahrd}. This is owed to Gaia's extremely precise astrometry, which enables the measurement of the best distances to stars performed so far via trigonometric parallaxes, as well as proper motions. The Gaia HRD is revealing new insights into e.g. star formation dynamics \cite{kounkel2022}, late-stage evolution of white dwarfs \cite{kilic2018}, and the contents of the solar neighborhood \cite{GCNSshort}

Though much can be derived from the HRD alone, even more information is encoded in large data sets collected with recent and new space telescopes like the \kepler~Space Telescope \cite{borucki2010} and the Transiting Exoplanet System Satellite (\tess; \cite{ricker2010}). These telescopes add additional dimensions beyond colour and magnitude: in their search for extrasolar planets, they are designed to measure the magnitude of hundreds of thousands of stars very precisely as a function of time. \kepler~alone has collected time series from roughly 530,000 stars with a 30-minute cadence, over a total of nine years of operations. For about half of those stars--observed during its 4-year primary operations phase--these time series span the entirety of that phase, with only short interruptions for spacecraft housekeeping.  While \kepler~focused on specific regions of the sky, TESS has observed $\sim$85\% of the entire sky during its initial two year primary mission. Now in its second extended mission, TESS is approaching coverage of the entire sky with extremely information-rich time series, which can be linked to fundamental physical processes in stars in ways scientists are only just starting to fully understand.


Studies of these data sets reveal that stars themselves show variability on all timescales accessible to \kepler: acoustic oscillations on the timescales of minutes, hours-long flares, rotational signatures caused by starspots on timescales of days, as well as activity cycles that can take months to years \cite{nielsen2019}. They are related to a range of astrophysical processes: periodic dips in the time series reveal the presence of binary systems (or systems with a higher multiplicity; \cite{kirk2016}), a wealth of periodic signals can be related to stellar rotation as well as global oscillation modes of stars \cite{aerts2021}, stellar flares and stochastic variations in magnitude are generated by magnetic reconnection and by convection processes \cite{davenport2019}. Many of these processes are directly related to the global properties of the star like its age and its mass. 

The variability information contained in the time series of stars observed with \kepler~and \tess~is complementary to the information displayed in the HRD, suggesting that a joint display of this information might lead to new insights. Combining these data sets visually, however, is challenging: since the HRD is already so rich in information, adding additional channels of information requires a higher dimensionality than the HRD (and human cognition) can easily support, thus potentially making the resulting visualization too confusing to understand. 
In addition, astronomy has a long history of serendipitous discoveries: every time a telescope opens up a new view of the sky, we find something unexpected that drives our understanding of the universe. However, for hundreds of thousands to billions of objects, exploratory data analysis becomes an extremely difficult task, because we can no longer easily look at every single star we have observed individually.

Here \textit{sonification}--the process of translating data into quantities that can  be turned into sound, rather than images--  can provide an interesting alternative as a tool for science, outreach, and education alike. Astronomy and sound have been linked since ancient times, when the Greek mathematician Pythagoras suggested that celestial bodies create music \cite{rackham1967natural}. In more recent times, sonification has emerged as a viable alternative to more traditional visualization approaches in different scientific areas. While in sonification, observed measurements may be transformed into indirect quantities before turning them into sound, a special case is \textit{audification}, where measurements are directly translated into sound. There has been a proliferation of sonification work linking astronomy and music in a number of different ways \cite{tutchton2012,diaz2012,quinton2016sonifying,tomlinson2017solar,zcosmos_sonification,tuckerbrown2022}. Much like with data visualization, the implementation details of a sonification are guided by the goals and intended outcomes in creating them, usually one or a mix of outreach, scientific research and artistic project. These different goals have led to a range of different approaches.

\textbf{Sonification as art}: there have been a number of instances of astronomical data being used to create different forms of music \cite{winton2012sonification,droppelmann2018,rengel2018creating,vallelisa_2022}. Often used in outreach--e.g.~to illustrate the diversity of stellar systems hosting extrasolar planets\footnote{See e.g.~sonifications highlighted by the Office of Astronomy for Development of the International Astronomical Union \url{http://www.astro4dev.org/blog/2018/10/22/sonification-videos-and-web-apps/}}--these sonifications are aimed at harmony, sometimes at the expense of stripping out some scientific information in order to focus on specific aspects of the data. However, \cite{tomlinson2017solar} note that one of the strengths of sound is its capacity to elicit emotions in the listener, and thus careful sound design to bring out a specific scientific concept can aid engagement and learning in the contexts of outreach and teaching. 

\textbf{Sonification used in research:} astronomers have generated sonifications of astronomical data sets in a limited set of circumstances, often to allow for a different experience of their data sets. The focus of these sonifications is usually primarily on the added scientific value compared to its artistic merit \cite{tutchton2012,quinton2016sonifying,cooke2019}. Astronomical time series are a natural target for turning into sound, and there are a range of sonifications that translate brightness directly into pitch. \cite{cooke2019} point out the potential usefulness of sonification in the context of analyzing large, complex data sets, where sonification can provide additional channels of encoding information that would otherwise clutter a visualization. \cite{zcosmos_sonification} created a sonification of the zCOSMOS galaxy data set, translating a high-dimensional data set of galaxy properties into various types of sound. They, among others, particularly emphasize the use of the \textit{principle of ecological metaphors} whereby sonifications are generated in such a way that the relationship between sound and the underlying data matches the intuitive expectations of the listener (for example, by connecting sound volume to absolute magnitude of a galaxy). Recently, \cite{tuckerbrown2022} explored whether sonification could be used to identify exoplanet transit in time series using the \textit{astronify} tool\footnote{\url{https://astronify.readthedocs.io/}}, and found that both experts and non-experts perform well on high-quality data, whereas noisy data may require additional auditory training. 

\textbf{Sonifications as an alternative to visualizations}: Researchers have proposed sonifications as an alternative to data visualizations, especially for blind and vision-impaired (BVI) researchers and to make outreach more accessible. For example, \cite{garcia2019} point out that astronomical data sets and archives are largely inaccessible to BVI individuals and that this hinders BVI students and researchers from participating in scientific research in astrophysics. Similarly, both \cite{tomlinson2017solar} and \cite{harrison2022} lay out that planetariums are almost entirely vision-centric, and are thus inaccessible to BVI visitors. \cite{tomlinson2017solar} build a planetarium show that integrates both visual and auditory information with storytelling to engage a broad audience in astronomy and to convey knowledge about the solar system. More broadly, \cite{cooke2019} build on work in the cognitive sciences that suggests that certain data modalities are better suited to listening comprehension compared to vision--for example, tracking changes over time--and that thus certain data analysis tasks are better performed using sonifications of data rather than visualization. Finally, in certain decision-making scenarios that involve large amounts of information, sonification can enable efficient multi-tasking, where the researcher in question performs a listening talk (e.g.~listening for anomalies in a data stream) while leaving other senses free for parallel activities.

In this paper, we present a new sonification pipeline for \kepler~observations of stars, and sonify 1958 stars across the full range of stellar properties and ages. The sonifications are minimally processed so as to accurately reflect the information contained in them, and in particular to retain a meaningful representation of the differences in variability between stellar types. The sonifications clearly reveal the presence of binary stars, stellar rotations, and flares. We combine this information with the absolute magnitude, colour, and derived distances for these stars from the \gaia~Data Release 2 \cite{gaiadr2}. The result is a multi-dimensional and multi-modal catalogue of information that can be further enhanced and used for different purposes (e.g.~scientific research, outreach, artistic expressions). Here, we combine our new tool and pipeline with a specific example designed for outreach and teaching: an interactive HRD, realized as a public website, where the HRD can be explored visually and sonically. This new website enables an exploration of the information contained within stellar variability in relation to the visual (positional) information contained within the HRD: how does stellar variability change as a function of position on the main sequence? Or as a function of position on the giant branch? Information contained in the sonified representations can be directly related to the physics of stellar structure and stellar dynamics, and thus provides a fuller picture of the underlying processes, while at the same time engaging learners through the novelty of sonification.

In Section \ref{sec:dataprep}, we introduce the astronomical data sets used in this work. In Section \ref{sec:sonification}, we introduce sonification in more detail and lay out the processes and pipelines we built to turn \kepler~time series into sound. Section \ref{sec:sonifiedhrd} introduces the sonified HRD, its current implementation as a public, interactive website (Section \ref{sec:websitedesign}), and highlights a number of illustrative example stars and their sonified time series (Section \ref{sec:examples}). Finally, in Section \ref{sec:discussion} we highlight current limitations and future directions of this work.

\section{Data Preparation}
\label{sec:dataprep}

We base our sample selection on the crossmatch between the \kepler~Stellar Table \cite{batalha2010} and \gaia~Data Release (DR) 2 \cite{gaiadr2} database provided by \cite{bedell2019} with a 20 arcsecond search radius in order to identify stars in both catalogues that closely match in position on the sky. We used \gaia~$G_{BP}$ and $G_{RP}$ magnitudes, photometric colour $G_{BP} - G_{RP}$ and parallax-derived distances from \cite{bailerjones2018} to construct a standard HRD. For sonification, we focus on data from the original \kepler~survey \cite{borucki2010} in order to maximize the total baseline of time series data available for sonification. Each quarter of three months of data translates into only a few seconds of sound, thus our goal was to select targets with as much uninterrupted time series data as possible.  

The aim of this project is to showcase the breadth and diversity of stellar variability, rather than present the HRD of an existing stellar population. With this in mind, we subsampled the catalogue using a reweighting scheme in order to oversample stellar populations that are comparatively rare (late-type stars, giants, white dwarfs) and undersample common stellar types (most main sequence stars). In addition, we hand-selected 17 sources of special interest into a ``curated'' set of sources to be highlighted and described in more detail (two of these sources are discussed in Section \ref{sec:examples}). 
For each source, we use the \textit{lightkurve} \textit{python} package \cite{lightkurve} to download and process the associated \kepler~light curves. Here, we focus exclusively on \kepler~long-cadence data with a time resolution of 30 minutes such that the sonifications of all stars are directly comparable to each other, and audible differences in sound are due to differences in the time series rather than changes in measurement cadence. We follow standard data processing procedures recommended by \textit{lightkurve} on the presearch data conditioning  simple aperture photometry (PDCSAP) flux. Because we used the long-cadence data, almost all of our time series consist of 16 quarters of observations. We used the default quality bitmask to ignore data segments with severe quality issues. Using \textit{lightkurve}, we stitched together multiple quarters, removed NaN values and removed systematic trends in the time series using a Savitzky-Golay filter with a window length of 801 samples. In total, our subsampled HRD contains 1958 sources. 

\section{Sonification}
\label{sec:sonification}
Data from \kepler~time series is sonified through a direct synthesis process in which a sound sample is synthesized for each data point.\footnote{ This process is similar to "audification", a sonification technique defined by Florian Dombois and Gerhard Eckel as: ``...a technique of making sense of data by interpreting any kind of one-dimensional signal (or of a two-dimensional signal-like data set) as amplitude over time and playing it back on a loudspeaker for the purpose of listening''. \cite{hermann2011sonification}, Chapter 12} 
As a result of this process we obtain sounds that have identical properties to the original time series of the stars but in a different timescale. The main parameter for this process in the digital domain is the sampling rate at which the data is synthesized. This parameter defines both how timescales in the original time series are translated into sound as well as the duration of the sonified light curve. Our goal is to preserve all the variability traits of the time series in the synthesized sound, allowing that periodic variability that happens at rates of cycles per day in the astronomical data is translated into a timescale in which they occur at rates of cycles per second so they could be perceived as sound. For example, brightness variability such as eclipses can be heard as rhythms (up to 20Hz), while faster types of periodic variability (above 20Hz) can be heard as pitch, or spectral components of the sound. Other more subtle variations in brightness can be perceived as slow amplitude modulation of the sound. After experimenting with different sampling rates we choose 24KHz to synthesize \kepler~long-cadence data. After the data pre-processing described in Section \ref{sec:dataprep}, further minimal processing is required before synthesis of the time series into sound waves that can be played through headphones or a loudspeaker.
This processing includes the following steps:

\textbf{Time interpolation:} data points of a \kepler~time series are time-stamped, and their times are corrected to the solar system barycenter such that they are not spaced uniformly. In order to represent the time series as sound digitally we resample them at uniform time intervals by means of linear interpolation using \texttt{scipy.interpolate.interp1d}.

\textbf{Filling data gaps:} Aside from long gaps in the data introduced by monthly data downlinks to Earth, shorter gaps appear in the time series data at irregular time intervals. While long data gaps are translated into silence in the sonification, short gaps produce popping sounds, thus we fill these gaps by means of one-dimensional interpolation to mitigate this problem.

\textbf{Median shift and normalization:} in order to create sound pressure waves we need the data to have positive and negative values that produce the corresponding compression and rarefaction of the air when reproduced. Therefore, we shift all the data values by the mean of the light curve. During this process the data is also normalized. \\

Once conditioned, we synthesize the data samples using 32-bit floating point resolution and a sampling rate of 24KHz and store the output into a sound file with WAV format. This approach to sonification has multiple advantages, the most important one being that no data is lost, which allows us to further analyze these time series as sounds. For instance, we could in practice track deterministic (tonal) components in the sounds and separate them from the stochastic (noisy) components, and re-synthesize them as separate sounds. By doing this we can further analyze the structure of the deterministic components which represent periodic changes (such as rotations, eclipses, etc.) and look closer at the structure of the stochastic components, which represent rapid and chaotic changes in brightness (such as flares). Another advantage of using this methodology is that time series of astronomical objects are translated into sound objects\footnote{ A sound object was defined by Pierre Schaeffer as: "This unit of sound [sound-object] is the equivalent to a unit of breath or articulation, a unit of instrumental gesture. The sound object is therefore an acoustic action and intention of listening". \cite{pierre1966traite}, page 271; English translation in \cite{schaeffer2012search}}. 
In other words, as they have a short duration, sonified time series can be aurally perceived in a way akin to the way their visualizations are visually perceived: as one gestalt. Most of the time series of the stars we selected have 4 years of data which were compressed into about 6 seconds of sound. The code to reproduce these sonifications, the sub-selected Gaia-Kepler cross-matched catalogue, and the data set of sonifications for all stars are available on \href{https://zenodo.org/record/7468367}{Zenodo}.

\section{The Sonified HR Diagram}
\label{sec:sonifiedhrd}

We combined relevant pieces of the original cross-matched data, including \gaia~magnitudes, \kepler~Input Catalogue (KIC) numbers, processed time series and sonifications into a data base for further development. The resulting multi-dimensional and multi-modal data set forms the basis of a wide range of possible modes of interaction. For this initial project, we chose an interactive website\footnote{accessible under \url{https://starsounder.space}} as a means to provide a means for both researchers and the public to interact with the data. 

\begin{figure}
    \centering
    \includegraphics[width=0.5\textwidth]{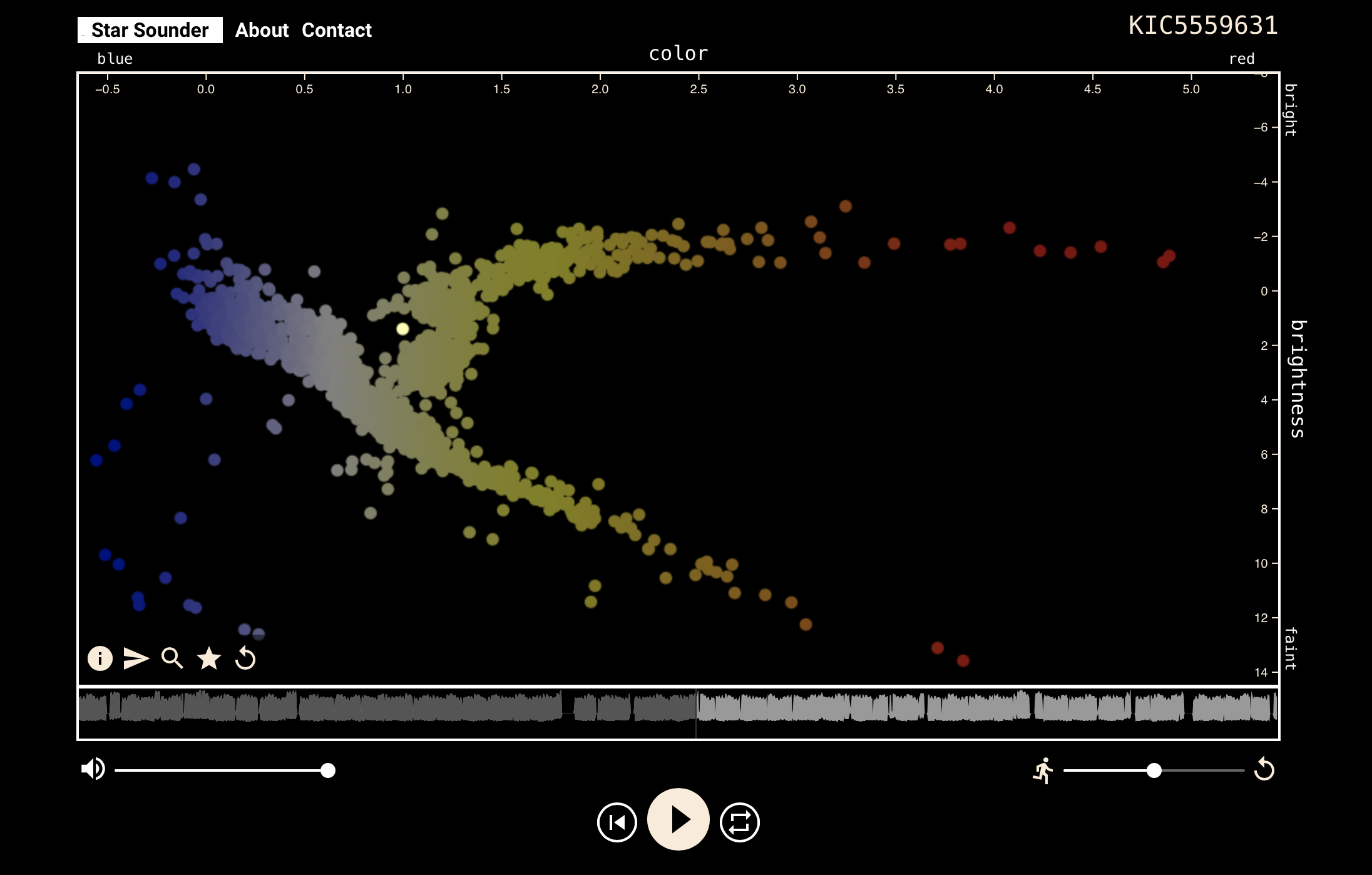}
    \caption{A screenshot from the interactive \url{https://starsounder.space} website. We show a visual representation of the classical HRD, with the currently selected star highlighted in intensity and its KIC identifier presented in the upper right. A representation of the time series is shown in grey below the diagram. Sound can be toggled via the large ``play'' button, additional controls enable changing volume and speed. Additional controls include an information panel, a ``share'' button generating a link, a search field, controls to toggle different stellar types, and a ``reset'' button.}
    \label{fig:website}
\end{figure}

\subsection{Designing an Interactive Website}
\label{sec:websitedesign}

Our goal is to combine data visualization, and in particular, a data visualization fundamental to stellar astronomy and familiar to researchers, with additional variability information encoded in auditory channels, to provide an enhanced representation of stellar properties (for a screenshot of the website, see Figure \ref{fig:website}). To do so, we start with a classical HRD, visualized as a scatter plot of \gaia~ absolute magnitudes against \gaia~ $G_{BP} - G_{RP}$ colour of our entire catalogue. We use colour gradients as a means of broadly encoding stellar colour. We note that this choice can be confusing to non-experts, since stellar colour is a proxy for temperature. The relationship between stellar colour and temperature (bluer stars are hotter, redder stars are cooler) defy every-day expectations of the link between colour and temperature, where the encodings are often reversed. 

We employ common language rather than astronomical jargon (e.g.~``brightness'' instead of ``magnitude'') in axis labels to aid non-experts in understanding the diagram, and provide additional information on the trends of the axes (e.g.~ faint versus bright at the lower and upper ends of the y-axis) for clarity. Basic interactivity in the HRD includes zooming and panning. Hovering the mouse pointer over a point in the HRD will present a floating box with the source's KIC number, to help experts rapidly identify sources of interest.

In the visualization of the HRD alone, variability is not immediately apparent, but interactivity provides access to these additional dimensions of information. Clicking on a single point in the HRD (corresponding to a source) highlights that point in the HRD by increasing its intensity, but also queries the database for the corresponding sonification. The sonification is displayed visually as a time series below the HRD, and is also accessible in its sound version through the ``play'' button prominently displayed below the HRD. Users can further interact with these sonifications in various ways. For example, clicking on the visual display of the sonification starts the playback of sound at this time stamp. Because the dynamic range of the sonifications is quite large, owing to the variety in variability amplitudes across different types of sources, we include a slider that can be used to adjust the sonification volume. A second slider enables the user to adjust the sonification speed: because different physical processes operate at different timescales, changing the speed can enable a richer understanding of the different components that comprise the sound the user hears.

To augment the user experience, we include a number of panels that can be opened and closed interactively, and provide additional controls and information. These include an information panel, a copy button, a search bar, a filter tool and a reset button. The information panel presents basic properties for the currently highlighted source (colour, magnitude, temperature and mass), compiled from the \gaia~DR2 catalogue. For curated sources, this panel also includes a text description of the source itself, including the type of source (e.g.~an eclipsing binary source), background information on its variability, guidance for what can be discerned in the sonification, and what physical processes the sounds correspond to.

The copy button will simply copy the link to the website, with the current star selected, to enable easy sharing of sources and sonifications with others. The search bar enables users to type a KIC number in order to identify specific stars and is largely meant for experts to quickly find specific sources of interest. The star filters can be used to select and de-select certain types of sources based on their positions in the HRD, such as white dwarfs, red giants and main sequence star. Importantly, it is possible to filter out all but our curated set of 17 stars selected for their astrophysical interest and sound diversity: these stars have augmented descriptions, and can act as a guided tour through the sonified HRD. Finally, the reset button resets the sonification to its original state, removing all filters as well as any zooming and panning effects.

\subsection{Illustrative Examples}
\label{sec:examples}

Below, we present a small number of illustrative examples of particular stars in our sample with interesting time series properties that in turn translate into rich and interesting sonifications. To hear the corresponding sonifications, we encourage the reader to make use of the ``search'' interface on the website. This interface is accessible via the magnifying-glass button, and to find a sonification, one may type the KIC number (without the ``KIC'' prefix) corresponding to the star into that interface. 

\subsubsection{V1504 Cyg (KIC 7446357)}

V1504 Cyg is a cataclysmic variable (CV) of SU UMa subtype. It contains a white dwarf (WD) and a low-mass (K or M-type) star which transfers mass onto the WD via an accretion disk. CVs are a diverse class with various types of behaviour. In particular, they exhibit dwarf novae: regular outbursts where the brightness of the accretion disk dominates the system. 
Su UMa variables are known to exhibit \textit{superoutbursts}, around a factor of 5 brighter and 4-5 times longer than regular dwarf novae. 
\begin{figure}
    \centering
    \includegraphics[width=0.45\textwidth]{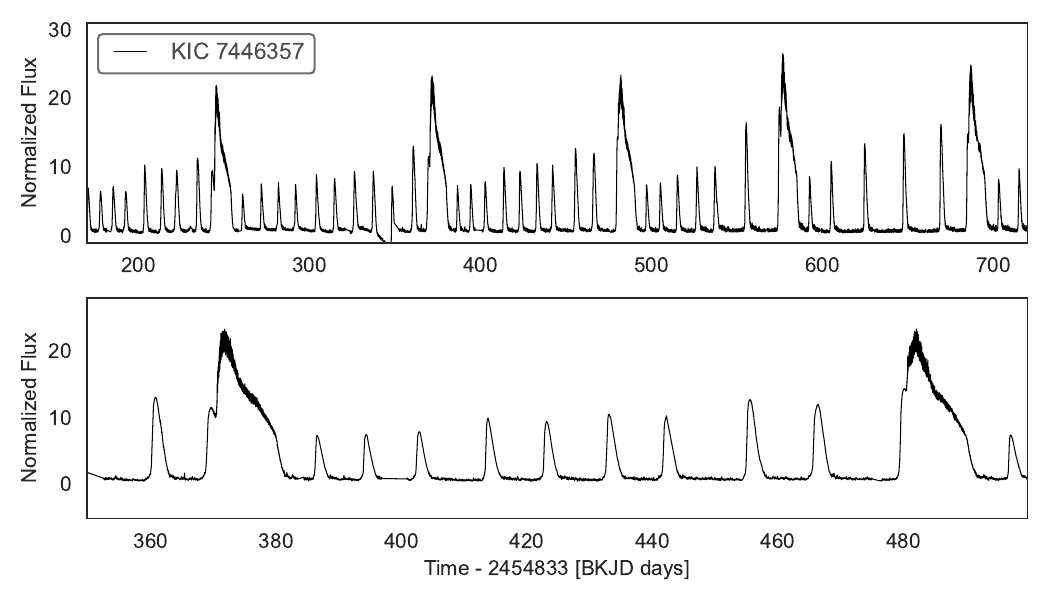}
    \caption{\kepler~time series for V1504 Cyg (KIC 7446357). The 16 quarter amount to a total of $\sim1421$ days, of which we show a segment of $550$ days (top) for illustration, and a zoom-in of $40$ days in the bottom. Clearly visible are both outbursts and superoutbursts occurring on timescales of slightly over ~1 month. }
    \label{fig:v1504cyg_lc}
\end{figure}

These outbursts and superoutbursts collectively display a rich temporal phenomenology (see Figure \ref{fig:v1504cyg_lc} for examples). Of note are the positive and negative superhumps seen during superoutbursts: these are hump-shaped distortions in the accretion disk that oscillate on a slightly longer (positive superhumps) or smaller (negative superhumps) period than the orbital period of the system. Positive superhumps are thought to be created by periodic compression of the accretion disk opposite the secondary star. In contrast, negative superhumps might be created when the accretion disk tilts out of the orbital plane, leading to the incoming accretion stream sweeping across the face of the disk, rather than its edge. Both positive and negative superhumps are observed simultaneously during superoutbursts. 

V1504 Cyg is one of sixteen CVs in the original field of view of the \kepler~mission. It was observed for 16 quarters and its variability behaviour has been well-studied, including outbursts, superoutbursts \cite{cannizzo2012} and stochastic variability \cite{dobrotka2016}. \cite{cannizzo2012} studied 459.8 days of \kepler~observations and showed that regular outbursts generally decay faster than exponential, with the largest deviance from an exponential decay close to outburst maximum. They studied four separate superoutbursts in the data, and shown that the burst duration of regular outbursts increases by a factor of $\sim\,1.2017$ between consecutive superoutbursts. 
V1405 Cyg was the only source in our sample that has also been target of a prior sonification study by \cite{tutchton2012}. The latter were particularly interested in the superhump behaviour, and thus pre-processed the time series to remove large-amplitude dwarf novae outbursts and superoutbursts using a high-pass box-car filter. The residual time series was folded to find average pulse shapes, yielding 30 distinct pulse shapes with frequencies determined by the observed superhump periods. In contrast to our work, these pulse shapes were then recombined into a single time series and directly rendered into a 30-second uncompressed sound file, encoding pulse profile brightness as pitch. In addition, they also directly render the time series of residuals into sound directly. They find that while each superoutburst is unique, there is an apparent repetition of tones throughout the superhumps, and this progression is unique to each of the two systems they study. There are also faint pulses in the noise that can be heard directly preceding each superoutbursts, rendering these events easily identifiable. 
\begin{figure}
    \centering
    \includegraphics[width=0.5\textwidth]{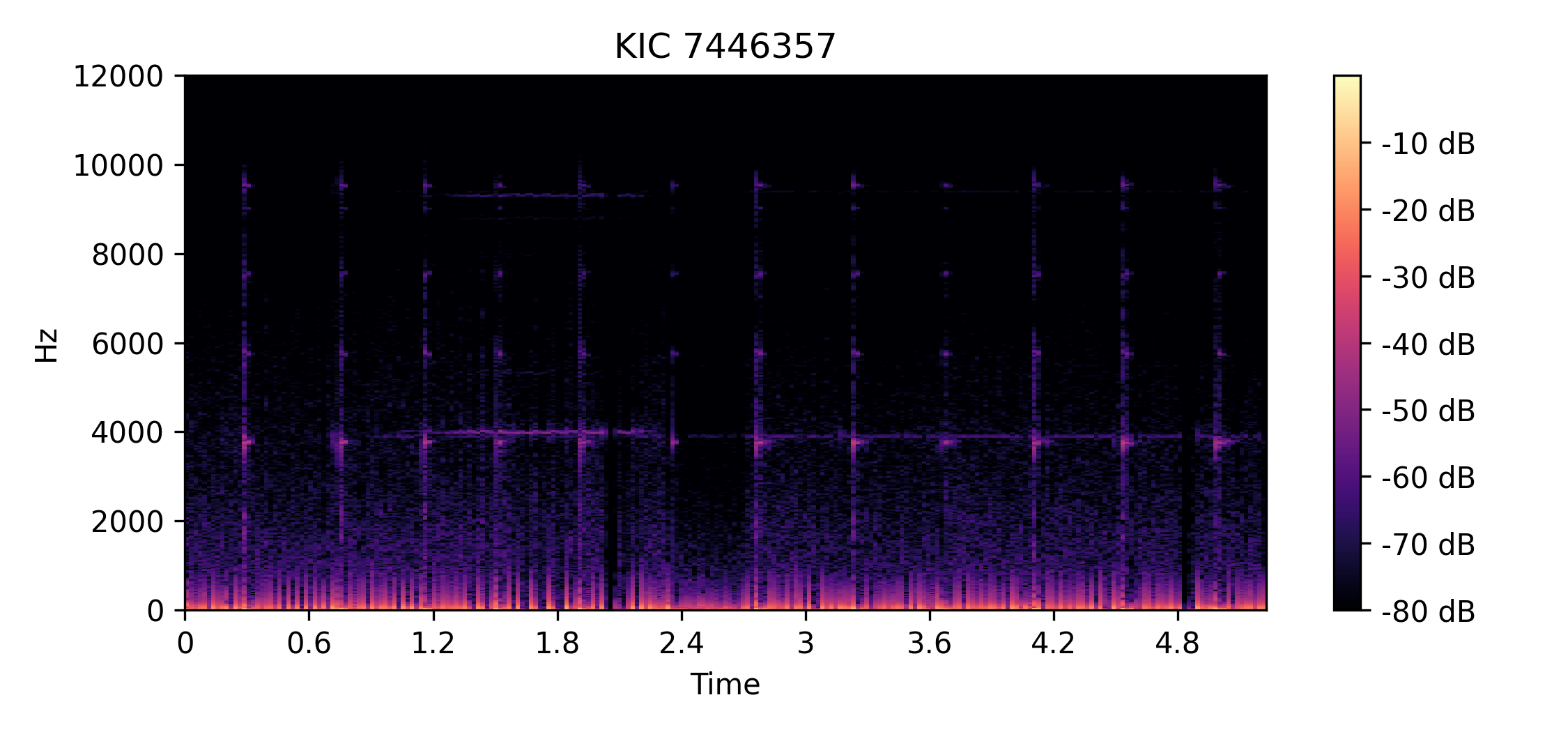}
    \caption{The sonification, and corresponding sonogram shown here, for V 1504 Cyg is extremely rich and has many layers (sound can be played on the corresponding \href{https://starsounder.space/?white+dwarfs=true&sub+giants=true&blue+giants=true&red+giants=true&main+sequence=true&curated=true&starID=7446357}{starsounder.space page}). The two most prominent layers are a sequence of high-frequency impulses (around 2 per second) which is produced by superhumps, and a sequence of fast rhythmic impulses (around 10 per second) which is produced by regular outbursts. The sonogram also shows a continuous partial of 3669 Hz frequency that starts at around 1 second and ends at around 2.4 seconds, and another sustained partial of 3464 Hz frequency that starts at around 3 seconds. These partials correspond to negative and positive superhump signals which occur when the accretion disk of the system tilts out of the orbital plane.}
    \label{fig:v1504cyg_sonogram}
\end{figure}
As can be expected from a system with such a complex variability phenomenology, our sonification of this binary system is extremely rich and has many layers of sound. The two most prominent layers are a sequence of high-pitched impulses (around 2 per second) which is produced by the superhumps, and a sequence of fast unpitched rhythmic impulses (around 10 per second) which is produced by the regular outbursts (both visible in Figure \ref{fig:v1504cyg_lc}). Listening more carefully one can also hear a high-pitched continuous tone starting at around 1 second and ending at around 2.5 seconds; there is also a fainter second sustained tone of lower pitch that starts at around 3 seconds. These tones are produced by negative superhump signals, difficult to see directly in Figure \ref{fig:v1504cyg_lc} without additional processing of the light curve, but readily apparent in the \textit{sonogram} (Figure \ref{fig:v1504cyg_sonogram}). The sonogram is a two-dimensional representation of power (or audio amplitude) as a function of time and frequency. Also called a \textit{spectrogram} in some contexts, it provides a visual representation of the sonification. 
\begin{figure}
    \centering
    \includegraphics[width=0.45\textwidth]{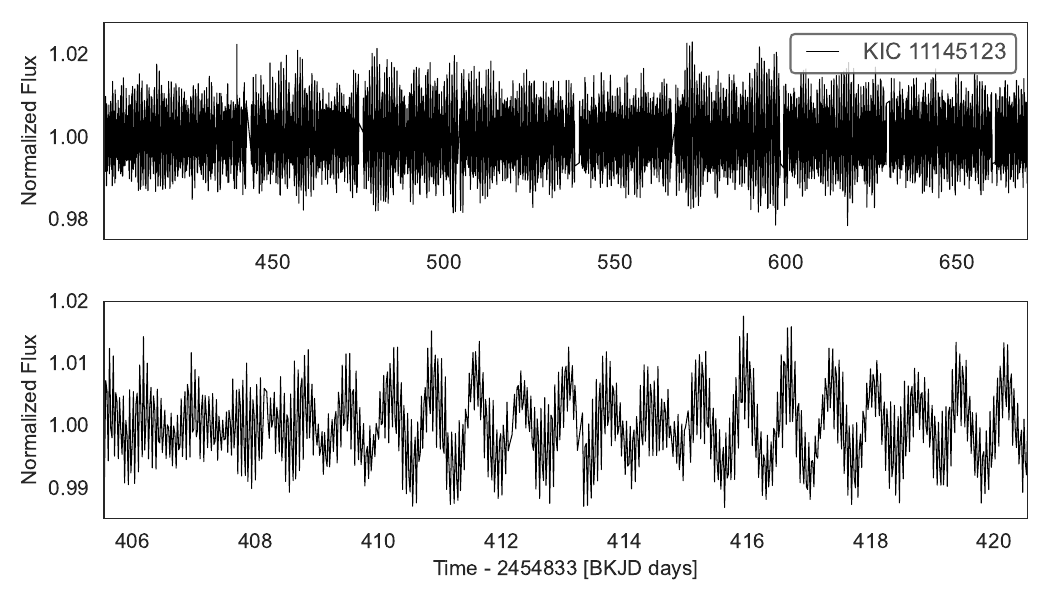}
    \caption{\kepler~time series for KIC 11145123. In the top panel, a segment of 270 days shows the overall noisy variability over long timescales. In contrast, shorter the botton segment reveals a range of quasi-periodic and periodic oscillations on multiple different time scales.}
    \label{fig:kic11145123_lc}
\end{figure}

\subsubsection{KIC 11145123}
KIC 11145123 is an example of a hybrid between two classes of pulsating stars: $\delta$ Scuti and $\gamma$ Doradus stars.
$\delta$ Sct stars show multi-periodic brightness variation, primarily high-frequency variations in the range of $\sim 0.008-0.42$ days with large amplitudes. 
These are thought to be both radial and non-radial pressure modes and can probe the structure and properties of the stellar envelope. $\gamma$ Dor stars, on the other hand, show low-frequency pulsations that are probably due to gravity modes closer to core regions that enables studies of the structure and composition of stellar cores.
Because the instability strip of both $\delta$ Sct and $\gamma$ Dor stars partially overlap in the Hertzsprung-Russell diagram, one might expect hybrids exhibiting the behaviour of both classes, i.e.~both high-frequency p-modes indicative of $\delta$ Sct behaviour, as well as the low-frequency g-modes of $\gamma$ Dor stars. These stars are particularly interesting, because the presence of both modes enables us to break degeneracies in stellar structure models present when modelling only one set of pulsations.

KIC 11145123 is a main-sequence A-type star of such hybrid type, showing both p-,g- as well as mixed modes, which have been used to infer the internal rotation profile and sphericity. The time series in Figure \ref{fig:kic11145123_lc} show noisy variability, but zooming in (Figure \ref{fig:kic11145123_lc}, bottom panel) clearly reveal periodic and quasi-periodic oscillations on multiple timescales. 
\cite{gizon2016} measured the sphericity of the star, and found its oblateness to be significantly smaller than expected even for a star with a slow rotation period of $\sim 100$ days, they called it the ``roundest object in the universe''. 

Our sonification reveals a symphony of stellar brightness variations expected from a $\delta$ Sct-$\gamma$ Dor hybrid star and from this particular star's complex time series. It is possible to hear three separate partial clusters, one with a center frequency at around 350 Hz (which sounds like a choir), a second one with a center frequency at around 4490 Hz and a final one with a center frequency around 7750 Hz, corresponding to the different g- and p-modes of the system, respectively. These partial clusters are also clearly visible in the sonogram in Figure \ref{fig:kic11145123_sonogram}.

\begin{figure}
    \centering
    \includegraphics[width=0.5\textwidth]{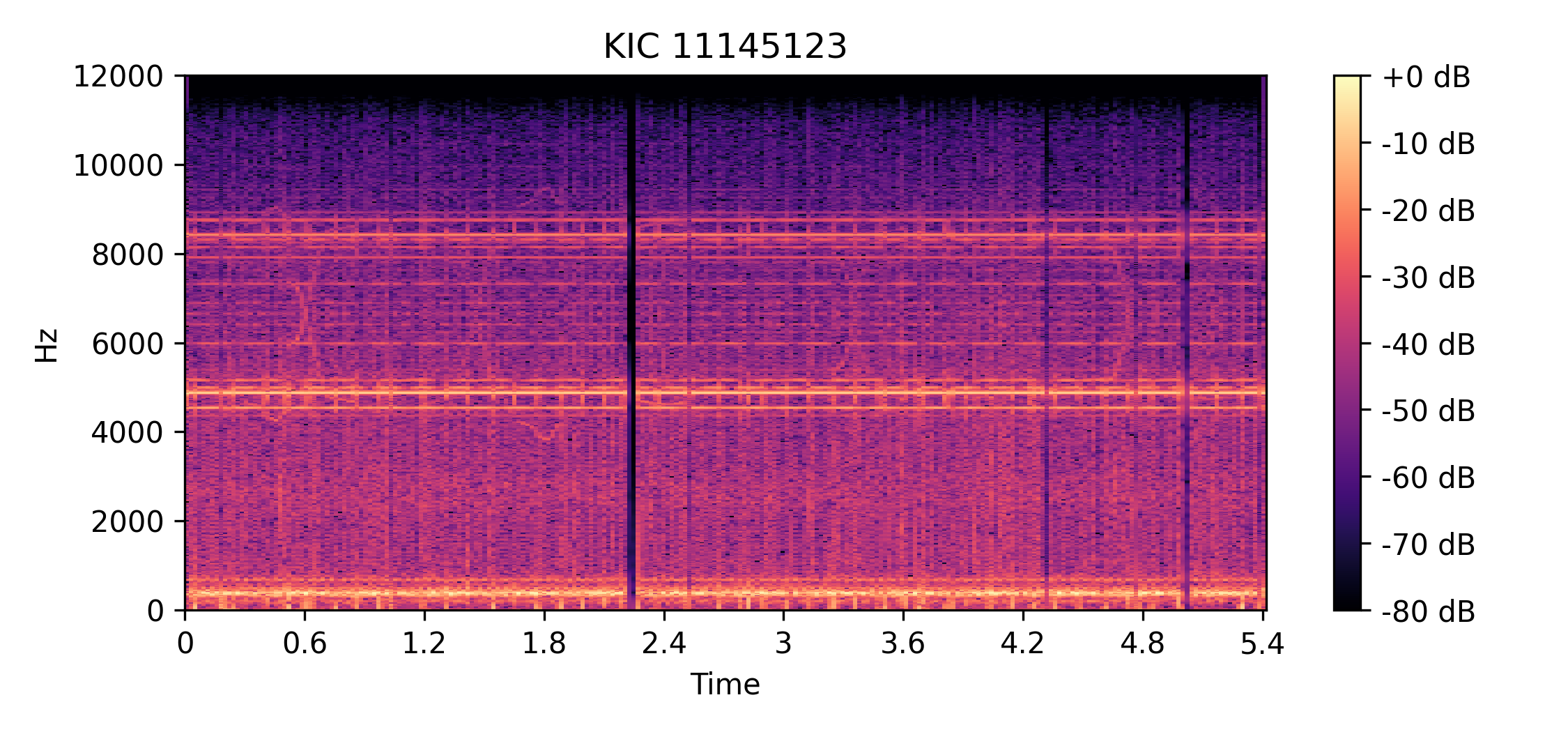}
    \caption{Sonogram of the sonification of KIC11145123. Three prominent clusters of partials can be seen, one with center frequency at around 350 Hz, a second one with center frequency at around 4490 Hz, and a last one with a center frequency around 7750 Hz. The corresponding sound is available on the star's \href{https://starsounder.space/?white+dwarfs=true&sub+giants=true&blue+giants=true&red+giants=true&main+sequence=true&curated=true&starID=11145123}{starsounder.space page})}
    \label{fig:kic11145123_sonogram}
\end{figure}

\section{Discussion}
\label{sec:discussion}
In recent years, sonification has become an increasingly popular tool for representing astronomical information. The emotional dimensions of sound can engage listeners from the general public and the scientific community alike \cite{tomlinson2017solar}. Sonification can both enable participation in research by BVI individuals, and also add additional dimensions of information to existing data representation. For some types of data, auditory processing may be superior to visual processing \cite{cooke2019}. 

\gaia, \kepler~and, more recently, \tess~have provided us with a wealth of new data of stars of all types and have enabled new perspectives on the astrophysical processes governing these stars, as well as new insights into long-standing astrophysical problems such as the Blazhko effect. Here, we have taken a large corpus of stellar time series observed with the \kepler~Space Telescope, and turned them into sound. We deliberately chose to minimally process the time series as one would do for scientific use, and directly translate key features into sound, without changes to the resulting sonification for aesthetic reasons. The key reason for this limitation is to enable direct comparison between different types of stars: how do high-mass stars sound differently from low-mass stars? How easy is it to recognize a binary system by sound? Our examples, chosen to represent a variety of types of stars and stellar variability, reveal the wealth of phenomena translated into sound: stellar rotation turned into constant pitch, the intricate binary interaction of outbursts and superoutbursts translated into amplitude variations of those sounds, and stellar flares as stochastic crackling.

Motivated by the fundamental nature of the Hertzsprung-Russell diagram in stellar research, we hypothesized that sonifications of stellar time series would add useful additional information to the classical HRD, without cluttering its visual representation. The HRD is a key diagnostic in exploring stellar structure and evolution, and by itself encodes a large amount of information about how stars form and evolve. We derived and visualized global stellar quantities from \gaia~observations, now supported by additional information provided by the \kepler~light curves, bringing together those data sets in a joint representation. Across the main sequence, as stellar mass and temperature change along with convective properties, changes in rotation period (pitch of a constant tone) and frequency of stellar flares (crackling) are apparent. 

Our decision process during the development of the sonification method was guided by (1) representing a range of astrophysically interesting time series properties into distinguishable sounds; (2) building sound representations where differences in sounds from different stars would meaningfully correspond to astrophysical differences (e.g. different rotation periods corresponding to different pitches) such that comparisons between stars are interesting and useful; (3) automating the process such that it can be applied to large catalogues of available light curves. Our approach has clear advantages in its scalability and interpretability of the resulting sonifications, but it also comes with limitations.
Importantly, the focus on scientific interpretability across a relatively broad range of time series features over aesthetics in the design process yields sonifications that are not always pleasant to hear. In teaching and outreach contexts, this might be detrimental to sustained engagement with the sonifications and the audio-visual HRD. 

Our bulk approach to data processing and cleaning with standard tools might lead to artifacts in the resulting input time series in individual cases. Indeed, the upper panel in Figure \ref{fig:v1504cyg_lc} shows a dip in flux at around ~$t_0 + 350$ d that is likely an artifact from data processing. While in this case, it has no effect on the global effects we aimed to sonify, \cite{plachy2014} have also shown that data processing techniques such as quarter stitching can lead to systematic effects on long-term amplitude modulations in Kepler time series. Thus, the results presented here are most useful for \textit{exploratory} studies, and precision astronomy with sonifications would likely require additional data processing steps hand-tuned to the source or class of sources at hand. 

One limitation of the current approach and interface lies in its serialized nature: a star is selected individually based on its position in the HRD, and the corresponding sound is played. This is of limited use in comparative analyses of stellar samples. Here, corpus-based concatenative synthesis \cite{schwarz2006real} might be an interesting extension of the current work: one extracts example sound units closest to a target sound from a large sound corpus, based on meta-data descriptors and features extracted from the sounds themselves. One could envision looking for clustering in a higher-dimensional space of stellar properties and sound features as a proxy of the underlying stellar variability, for example in a comparative study of specific types of stars such as RR Lyrae variables. 

We envision a range of different applications of our work. The website we have designed here may be useful in a teaching context. The HRD is a key component of introductory astronomy classes, and one could consider building exercises on top of the interactive representation presented here to engage students and deepen their understanding of stellar structure. Activities might involve a treasure hunt for stars based on sound, or a discussion of how the sound students hear relate to the underlying physics they have learned. 

While the HRD as represented here contains an interactive, visual component, the sonifications themselves do not, and indeed, much of the value of this project lies in the interdisciplinary combination of sonification practice with astronomical data and research. The HRD as a two-dimensional representation could be turned into an interactive touch display to remove the visual component. Similarly, \cite{tomlinson2017solar} and \cite{harrison2022} have shown that sonification adds a dimension to experiences that BVI individuals might otherwise be excluded from, such as planetarium shows. Our sonifications, available online on \href{https://zenodo.org/record/7468367}{Zenodo}, could help make these programmes more engaging for a wide range of visitors.  

\section{Conclusions}

In this paper, we have presented an automated pipeline to transform stellar time series into sound, and an interactive, audio-visual website that hosts these sound artifacts to enhance a well-known astronomical visualization, the Hertzsprung-Russell diagram. We show that our sonification method preserves astrophysically meaningful features of the underlying systems and data, and that the combination of a visual HRD derived from \gaia~data and the sonifications derived from \kepler~data enable a more in-depth exploration of stellar variability and structure than with either data sets individually.
The pipeline and the interactive HRD presented in this paper are the first result of a larger, overarching project exploring the use of sonification in the context of stellar astronomy. 
In the future, we will consider approaches that preserve the interesting structure while also provide a pleasant listening experience to users, and use user experience research and experiments to guide development of representations that combine pleasantness, user engagement and preservation of astrophysical information. While in this project, we focused largely on the use of sonifications of stellar time series in the context of astrophysical research and education, near-future work will include designing and implementing artistic concepts involving these sonifications.
Future work should consider comparative studies of the efficacy for learning of this new expanded version of the HRD compared to the traditional approach to teaching stellar evolution using the purely visual HRD. Development currently in progress includes combination of the sonification with virtual reality glasses in order to provide a more immersive experience, for example in a museum context. We are also exploring adjacent data sets of astronomical sources, including from the \tess~mission, as well as from compact objects. 

\section{Acknowledgments}

This research project was generously funded by the Bergstrom Award for Arts and Sciences of the College of Arts and Sciences, University of Washington. D.H. is supported by the Women In Science Excel (WISE) programme of the Netherlands Organisation for Scientific Research (NWO). We acknowledge support from the DiRAC Institute in the Department of Astronomy at the University of Washington. The DiRAC Institute is supported through generous gifts from the Charles and Lisa Simonyi Fund for Arts and Sciences, and the Washington Research Foundation. This work made use of the gaia-kepler.fun crossmatch database created by Megan Bedell. We are grateful to the eScience Institute at the University of Washington for facilitating the event that brought the authors together and started this project.

\bibliographystyle{IEEEtran}
\bibliography{hrdsonification}
%
%
%
%

\end{sloppy}
\end{document}